\documentclass[journal=jacsat,manuscript=article]{achemso}

\usepackage{amsmath}
\usepackage{amsfonts}
\usepackage{amssymb}
\usepackage{dcolumn}
\usepackage{epsfig}
\usepackage{graphicx}
\usepackage{latexsym}
\usepackage{setspace}
\usepackage{fancyhdr}
\usepackage{wrapfig}
\usepackage{ulem}
\usepackage{titlesec}
\usepackage{xcolor}
\usepackage{outlines}
\usepackage{multirow}
\usepackage{adjustbox}
\usepackage{enumitem}
\usepackage[official]{eurosym}
\usepackage{mdframed}
\usepackage{enumitem}
\usepackage{hyperref}

\usepackage[version=3]{mhchem} 

\author{Liao Duan}
\affiliation{Electrical and Computer Engineering Department, University of California, Santa Barbara, CA 93106, USA}
\author{Trevor J. Steiner}
\affiliation{Electrical and Computer Engineering Department, University of California, Santa Barbara, CA 93106, USA}
\alsoaffiliation{Materials Department, University of California, Santa Barbara, CA 93106, USA}
\author{Paolo Pintus}
\affiliation{Electrical and Computer Engineering Department, University of California, Santa Barbara, CA 93106, USA}
\alsoaffiliation{Physics Department, University of Cagliari, Monserrato, 09042, Italy}
\author{Lillian Thiel}
\affiliation{Electrical and Computer Engineering Department, University of California, Santa Barbara, CA 93106, USA}
\author{Joshua E. Castro}
\affiliation{Electrical and Computer Engineering Department, University of California, Santa Barbara, CA 93106, USA}
\author{John E. Bowers}
\affiliation{Electrical and Computer Engineering Department, University of California, Santa Barbara, CA 93106, USA}
\alsoaffiliation{Materials Department, University of California, Santa Barbara, CA 93106, USA}
\author{Galan Moody}
\affiliation{Electrical and Computer Engineering Department, University of California, Santa Barbara, CA 93106, USA}
\email{moody@ucsb.edu}

\title[Broadband entanglement generation]
  {Broadband Entangled-Photon Pair Generation with Integrated Photonics: Guidelines and A Materials Comparison}

\begin{document}

\begin{abstract}

\noindent Correlated photon-pair sources are key components for quantum computing, networking, and sensing applications. Integrated photonics has enabled chip-scale sources using nonlinear processes, producing high-rate entanglement with sub-100 microwatt power at telecom wavelengths. Many quantum systems operate in the visible or near-infrared ranges, necessitating broadband visible-telecom entangled-pair sources for connecting remote systems via entanglement swapping and teleportation. This study evaluates broadband entanglement generation through spontaneous four-wave mixing in various nonlinear integrated photonic materials, including silicon nitride, lithium niobate, aluminum gallium arsenide, indium gallium phosphide, and gallium nitride. We demonstrate how geometric dispersion engineering facilitates phase-matching for each platform and reveals unexpected results, such as robust designs to fabrication variations and a Type-1 cross-polarized phase-matching condition for III-V materials that expands the operational bandwidth. With experimentally attainable parameters, integrated photonic microresonators with optimized designs can achieve pair generation rates greater than ~1~THz/mW$^2$. 

\end{abstract}

\section{Introduction}

\begin{figure}[b!]
    \centering
    \includegraphics[width=1.0\columnwidth]{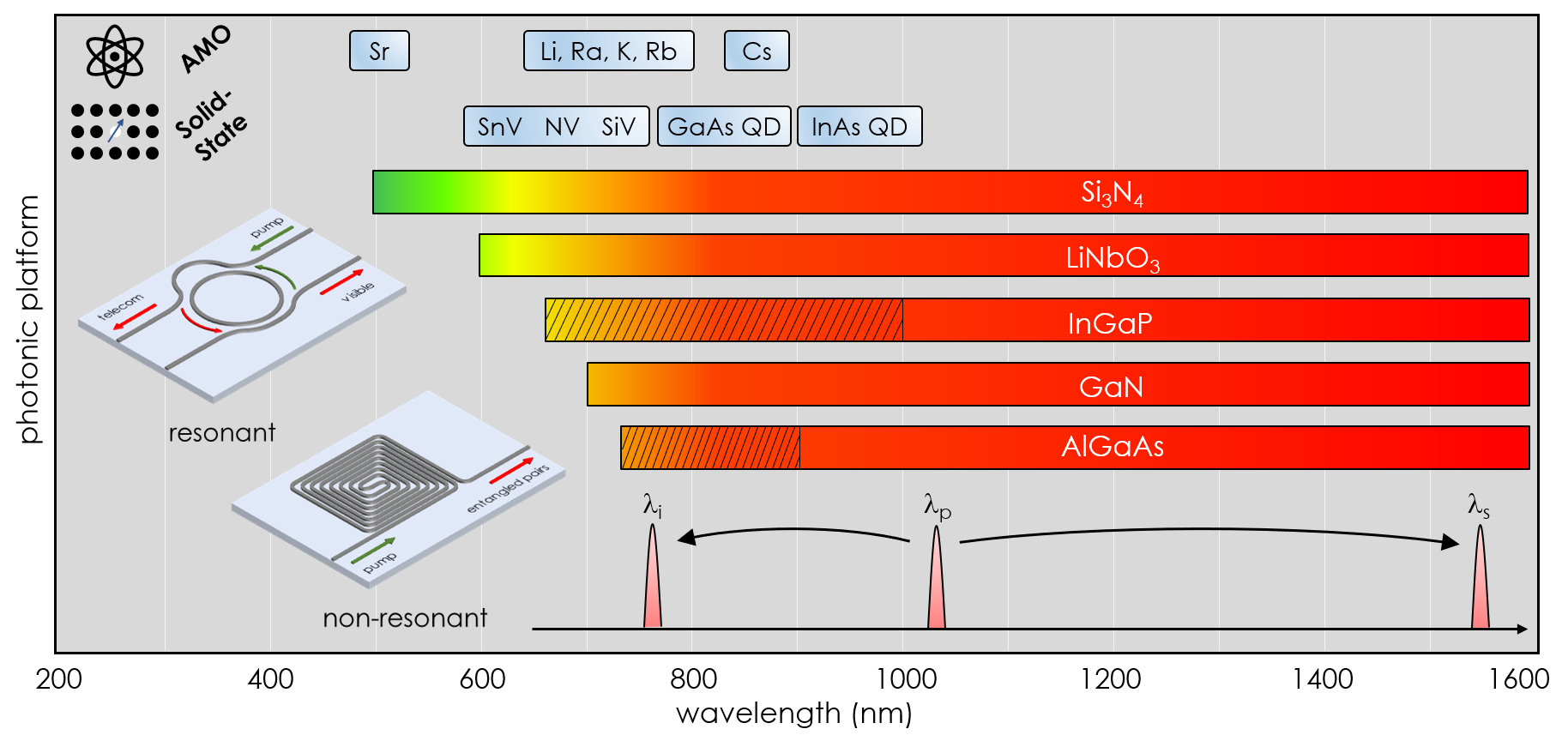}
    \caption{Comparison of the wavelength ranges for entangled photon pair generation across different material platforms. Entangled pairs can be generated at any two wavelengths within the range specified by the bar via waveguide geometry tailoring to achieve phase matching. For AlGaAs and InGaP, the shaded area indicates the extended wavelength range enabled by employing cross-polarized SFWM analogous to Type-1 spontaneous parametric down-conversion.\label{fig:intro}}
\end{figure}

Large-scale quantum networks and communication systems require a diverse set of components for generating, processing, storing, and transmitting quantum information with low loss and high fidelity~\cite{awschalom2021development,moody20222022}. For fiber-based networks, these requirements present unique challenges, as many quantum platforms that are being developed for quantum memories, repeaters, and processors, including color centers in solids~\cite{sukachev2017silicon}, trapped ions~\cite{wang2021single,pino2021demonstration}, neutral atoms~\cite{norcia2023midcircuit,buser2022single}, and semiconductor quantum dots~\cite{maring2024versatile,borregaard2020one}, operate in the visible and near-infrared wavelength range (Fig.~\ref{fig:intro}). Interconnecting such platforms in quantum network may be possible through the coherent conversion of emitted photons to telecom wavelengths, but quantum frequency conversion introduces new challenges in designing appropriate systems for specific wavelengths and filtering the strong optical pump required for efficient photon number conversion~\cite{singh2019quantum,zaske2012visible,bock2018high,morrison2021bright}. An alternative approach to bridge the visible-telecom wavelength gap is through broadband entanglement generation with suitably designed and engineered nonlinear optical platforms~\cite{lu2019chip}. In this scheme, remote quantum nodes can be entangled through successive entanglement swapping operations via a Bell-state measurement at a central node. 

Several types of visible-telecom bi-photon sources have been developed, including nonlinear optical fibers~\cite{soller2010bridging}, periodically poled bulk nonlinear crystals~\cite{clausen2014source}, and integrated photonic microresonators~\cite{schunk2015interfacing}. In crystals and fibers, pairs are generated over a broad bandwidth, requiring spectral filtering for entanglement swapping that also reduces the useful pair generation rate and efficiency. In the latter approach with integrated photonic resonators, incorporation of the nonlinear medium within the cavity enhances the entangled-pair generation efficiency. Using a stoichiometric Si$_3$N$_4$ microcavity, for example, Lu \textit{et al.} generated visible-telecom photon pairs between 668~nm and 1548~nm with high photon-pair generation rate (PGR), coincidence-to-accidental ratio (CAR), and two-photon visibility with sub-milliwatt optical pump power~\cite{lu2019chip}. The wide bandgap of Si$_3$N$_4$ makes it appealing for broadband pair generation; however, it also has a weak $\chi^{\left(3\right)}$ optical nonlinearity compared with other nonlinear integrated photonic materials~\cite{moody2020chip}. 

To this end, we examine different leading integrated photonic materials capable of visible-telecom entangled-photon pair generation, including silicon nitride (SiN), lithium niobate (LN), aluminum gallium arsenide (AlGaAs), indium gallium phosphide (InGaP), and gallium nitride (GaN). Using known dispersion curves for each material, we examine the optimal designs for phase-matched spontaneous four-wave mixing across the widest bandwidth spanning 1550~nm to the visible spectrum, covering several leading quantum memory and optical clock platforms such as quantum emitters in diamond and hexagon boron nitride (hBN), III-V quantum dots (QDs), and atomic systems including strontium (Sr), lithium (Li), radium (Ra), potassium (K), and rubidium (Rb). We then compare the performance of each platform in terms of entangled-pair brightness and bandwidth. We find that, of the materials examined, SiN provides the broadest wavelength separation of the signal-idler entangled photons due to its wide bandgap and flat dispersion, whereas III-V materials such as InGaP and AlGaAs offer the highest pair generation rates. Importantly, we find a novel Type-1 cross-polarized phase-matching scheme for the InGaP and AlGaAs, extending the idler photon further into the visible regime as indicated by the shaded regions in Fig.~\ref{fig:intro}. By analyzing the phase-matching conditions, we find optimal designs that simultaneously maximize the pair-generation rates and are highly tolerant to fabrication process variations, making these designs robust, scalable, and efficient for entanglement distribution.

\section{Methods}
\subsection{Material Properties}

When designing and configuring integrated photonic broadband entangled-photon pair sources, different semiconductor materials offer different advantages. Platforms like SiN are compatible with foundry manufacturing, facilitating the use of process design kits, mature fabrication and processing, and a wide bandgap that extends the accessible wavelength range for photon-pair generation. Additionally, SiN, known for its moderate third-order nonlinearity, can also achieve low-loss waveguides that can enhance the generation efficiency and allow for low-loss passive components to be created on chip. Materials such as AlGaAs, InGaP, and GaN excel in pair generation efficiency through spontaneous four wave mixing (SFWM) due to large third-order nonlinear coefficients and high refractive index contrast, but these platforms are not yet scalable at foundry levels and were historically limited by high propagation losses. Recent advancements have significantly reduced these losses to levels comparable to state-of-the-art silicon waveguides at telecom wavelengths around 1550~nm, making these platforms the brightest for pair generation at 1550~nm via SFWM and spontaneous parametric down conversion (SPDC)~\cite{akin2024ingap,kues2019quantum,zeng2024quantum,thiel2024}.

\begin{figure}[t!]
    \centering
    \includegraphics[width=0.5\linewidth]{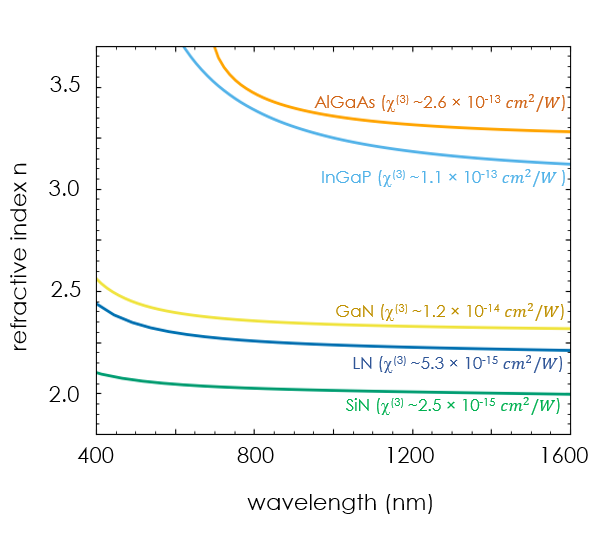}
    \caption{\label{fig:index}Refractive index of AlGaAs, InGaP, GaN, LN and SiN. As shown here, AlGaAs and InGaP have large slope near the visible spectrum, while GaN, LN, and SiN have an approximately linear slope across the wavelengths studied.}
\end{figure}

Entangled photon pairs are typically produced through nonlinear optical processes: SPDC in  $\chi^{(2)}$ materials and SFWM in $\chi^{(3)}$ materials. SPDC involves the annihilation of a single pump photon to create a photon pair, while SFWM involves two pump photons interacting to produce the pair, both adhering to energy conservation laws with $\omega_p=\omega_s + \omega_i$ for SPDC and $2\omega_p=\omega_s + \omega_i$ for SFWM~\cite{boyd2008nonlinear,helt2012does}. For generating entangled-photon pairs across visible and telecom wavelengths, SFWM is preferred because it uses more accessible pump wavelengths near 1,000~nm, unlike the 300-700~nm required for SPDC-based sources. 

In addition to energy conservation that specifies the frequency of generated photons, momentum should also be conserved. This condition is described by the wavevector mismatch formula
\begin{equation}\label{Eq:1}
    \Delta k = k_i + k_s - 2k_p,
\end{equation}
where $k=2\pi n_{eff}/\lambda$, $n_{eff}$ is the effective refractive index, and $\Delta k = 0$ ensures momentum conservation. Achieving the phase-matching condition is challenging due to material and waveguide dispersion, which cause the effective index to vary with wavelength. Rewriting this condition gives 
\begin{equation}\label{Eq:2}
\omega_s n_s + \omega_i n_i = 2 \omega_p n_p,
\end{equation}
where $n_s$, $n_i$, and $n_p$ are the values of the effective indices at angular frequencies $\omega_s$, $\omega_i$, and $\omega_p$, respectively. 

As a representative demonstration, we model different material platforms by pumping at 1033.3~nm and generating the entangled pairs at 1550~nm and 775~nm, motivated in part by near-IR transitions of the hyperfine levels of potassium and rubidium~\cite{gupta1973hyperfine}. Different wavelengths can also be chosen to connect different systems operating at shorter wavelengths. To achieve zero phase mismatch, following Eq.~\ref{Eq:2}, we must satisfy $2n_s + n_i = 3n_p$. Two routes are possible: search for a material system and geometry that yields identical values for  $n_i$, $n_s$, and $n_p$, or manipulate the waveguide geometry for a given material system to achieve a quasi-linear dependence on the effective index such that the increase in effective index of the idler mode is offset by a corresponding reduction in the effective index of the signal mode. Intuitively, from Fig.~\ref{fig:index}, GaN, SiN, and LN have flat and nearly linear dispersion in the relevant wavelength range, which simplifies achieving phase matching. As a result, assuming a signal wavelength of 1550~nm, the minimum idler wavelength for GaN, LN and SiN, is 700~nm, 600~nm, and 500~nm, respectively. 

Achieving phase matching becomes challenging when the idler photon wavelength approaches the electronic band gap of the material. For materials like Al$_{0.3}$Ga$_{0.7}$As and In$_{0.48}$Ga$_{0.52}$P (denoted simply as AlGaAs and InGaP below), with band-edge transitions near 700~nm and 650~nm, respectively, the nonlinear increase in the refractive index hinders broadband phase matching for visible-wavelength idler photons as shown in Fig.~\ref{fig:index}; however, because AlGaAs and InGaP are both cubic crystals with $F\bar{4}3m$ symmetry~\cite{yariv1983optical,Sadao1985}, their $\chi^{\left(3\right)}_{ijkl}$ tensors have non-vanishing elements for $\left[k=l,i=j\right]$ and $\left[i=k,j=l\right]$, where the subscripts $i$, $j$, $k$, and $l$ denote the signal, idler, and two pump fields for the SFWM process, respectively. Consequently, instead of restricting the pump, signal, and idler photons to all have the same polarization, which we denote as Type-0 pair generation (following the conventions for SPDC~\cite{boyd2008nonlinear}), a cross-polarized pump-pair polarization scheme can also be utilized, which we denote as Type-1. As shown in Table \ref{table:index}, for Type-1 pair generation, the non-zero element $\chi^{\left(3\right)}_{xxyy}$ allows for pumping the TM mode to generate signal-idler pairs in the TE mode. The $\chi^{\left(3\right)}_{xxyy}$ values used for Type-1 generation are taken to be half of $\chi^{\left(3\right)}_{xxxx}$ as an approximation~\cite{Hutchings}. This cross-polarized scheme can be leveraged to achieve phase-matching for large signal-idler wavelength separation, since the effective indices of the TM pump and TE pairs can be independently engineered through the waveguide geometry. In the numerical simulations in the following section, we explore pair generation for both Type-0 (for GaN, SiN, and LN) and Type-1 (for AlGaAs and InGaP) schemes.

\begin{table}[h!]
{\small
\centering
\begin{tabular}{|c|c|c|c|c|}
\hline
       & Pump \& Pump& Signal \& Idler & AlGaAs       & InGaP        \\ \hline
Type 0 & TE + TE   & TE + TE         & $\chi^{(3)}_{xxxx}$& $\chi^{(3)}_{xxxx}$ \\ \hline
Type 1 & TM + TM   & TE + TE         & $\chi^{(3)}_{xyxy}, \chi^{(3)}_{xxyy}$ & $\chi^{(3)}_{xyxy}, \chi^{(3)}_{xxyy}$ \\ \hline
\end{tabular}
\caption{\label{table:index} \small Relevant polarization schemes for Type-0 (co-polarized pump and pairs) and Type-1 (cross-polarized pump and pairs) pair generation via SFWM that are permitted by the non-vanishing tensor values of $\chi^{(3)}$.}}
\end{table}

\subsection{Simulations}

In the previous section, the design of broadband entangled photon pair sources using SFWM in nonlinear materials was introduced, focusing on achieving phase matching between the pump, signal, and idler modes. Here, we utilize the Ansys Lumerical finite difference eigenmode solver to tailor waveguide geometries for phase matching at the specified wavelengths. We aim for the signal photon to be within the telecom C-band at 1550~nm to minimize optical fiber propagation loss for long-distance transmission. In our approach, depicted in Fig.~\ref{fig:intro}, we sweep the idler wavelength starting from 1550~nm to determine the point where geometry modifications cannot achieve the zero frequency mismatch condition. The pump angular frequency, determined by $\omega_p=(\omega_i+\omega_s)/2$, varies as the idler wavelength is swept, and the signal wavelength remains fixed at 1550~nm.

\begin{figure}[t!]
    \centering
    \includegraphics[width=0.99\columnwidth]{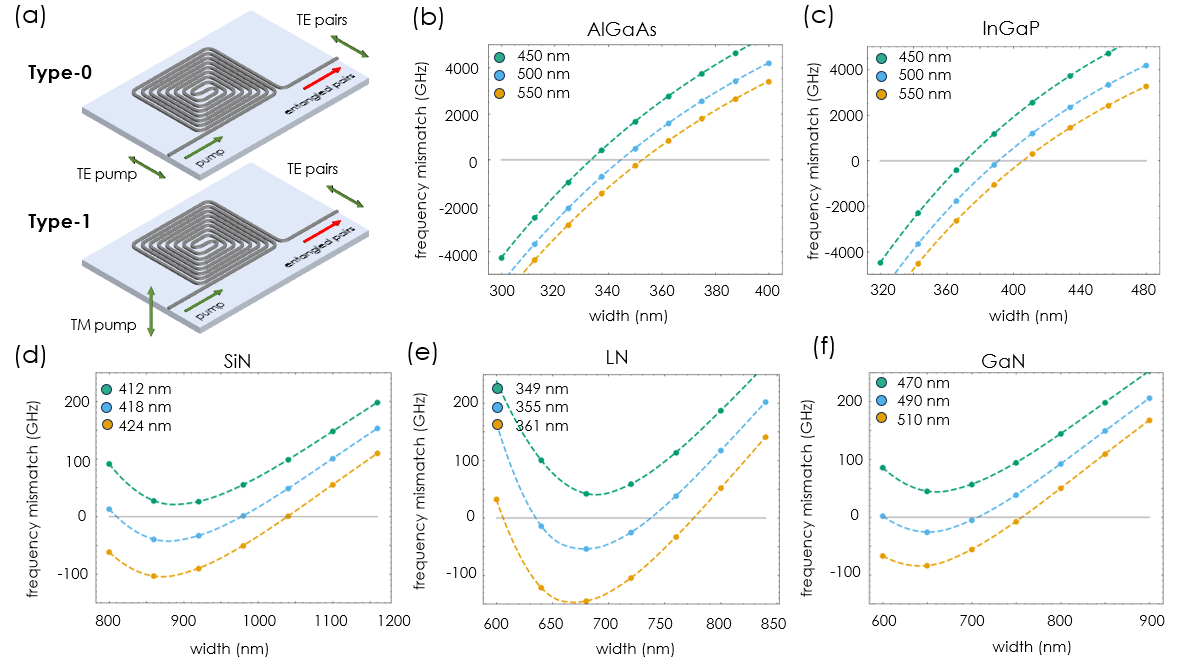}
    \caption{\label{fig:Simulation}Frequency mismatch simulations for different waveguide dimensions assuming linear waveguides. For ring resonator geometries, the simulations are modified to account for the change in refractive index for a given bend radius. Type-0 phase-matching is used for SiN, GaN and LN, while Type-1 is used for AlGaAs and InGaP.}
\end{figure}

As a specific case study, we model phase matching in these materials with an idler photon at 775~nm and a signal photon at 1550~nm, setting the pump wavelength at 1033.3 nm for energy conservation. The resulting frequency mismatch, $\Delta w = \frac{w_s n_s + w_i n_i}{2 n_p} - w_p$, is analyzed for various waveguide geometries shown in Fig.~\ref{fig:Simulation}. In this model, we employ the Type-0 phase matching for SiN, LN, and GaN, leveraging their relatively flat refractive index curves. For AlGaAs and InGaP, which exhibit a significant refractive index variation across the selected wavelengths, we use Type-1 phase matching. In our simulations for AlGaAs and InGaP, we set the waveguide dimensions near 350~nm by 500~nm, favoring TM mode confinement for the pump mode. Due to the reduced confinement of the TE mode using this structure at visible and telecom wavelengths, the mode extends into the SiO$_2$ cladding, thus lowering the effective index at these wavelengths. Note that for SiN and LN, the waveguides have only bottom SiO$_2$ cladding with a top air-cladding to increase mode confinement in the waveguide because these materials have relatively low refractive index contrast. Since the nonlinear interaction only occurs within the waveguide, weak mode confinement would be detrimental to the performance of these low-index contrast platforms. For the other material systems which have considerably larger refractive index contrast, the SiO$_2$ cladding surrounds the entire waveguide.

\section{Results and Discussion}

Simulations for SiN, GaN, and LN suggest that zero frequency mismatch can be achieved through Type-0 generation, provided waveguide dimensions are optimized. For those materials, their phase matching trends consist of two regimes. In the first regime, the trend is flat and the change in frequency mismatch remains small within a certain waveguide width. In the second regime, as the waveguide width is further increased or decreased, the frequency mismatch increases linearly with width. Interestingly, this behavior indicates that for SiN, GaN and LN, a unique thickness exists that permits $\leq10$ GHz frequency mismatch over $\geq100$ nm variation in waveguide width. Because fabrication errors more severely affect the waveguide width, the first regime is an optimal design point for minimizing the impact of these errors. For SiN, this specific height is 416 nm and the frequency mismatch is within $\pm$10 GHz with widths ranging from 800~nm and 950~nm. Similar results are found for LN and GaN, which are shown in Table~\ref{table:index2}. 

\begin{table}[h!]
\centering
{\small
\begin{tabular}{|l|c|c|c|}
\hline
 & {Height (nm)} & {Width (nm)} & {Frequency Mismatch (GHz)}\\ \hline
{SiN} & 416 & 800-950 & $\leq$10 \\ \hline
{LN} & 352& 630-730 & $\leq$15 \\ \hline
{GaN} & 485& 600-700 & $\leq$5 \\ \hline
\end{tabular}
\caption{\label{table:index2} \small Waveguide cross-section values for each material platform to achieve optimal pair generation with minimal frequency mismatch and fabrication process variation tolerance in the waveguide widths for fixed waveguide heights.}}
\end{table}

\begin{figure}[b!]
    \centering
    \includegraphics[width=0.99\columnwidth]{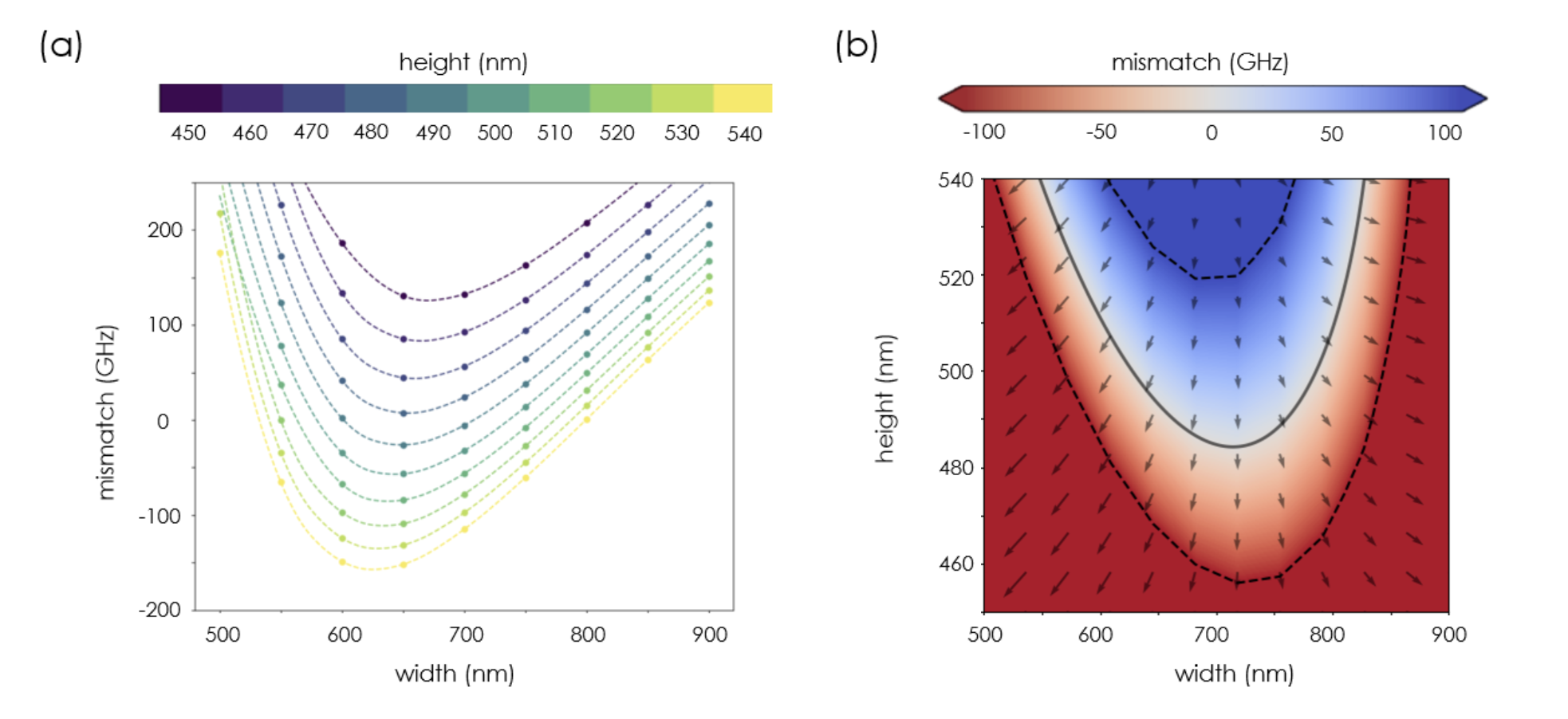}
    \caption{\label{fig: GaN_VectorField}  A more complete phase matching profile for GaN is shown in (a) with a wider range of height and width (b) heatmap as a combination of vector field and mismatch. The dashed contour lines show when the mismatch value is at $\pm$ 100 GHz. The solid black line indicates dimensions for zero frequency mismatch. }
\end{figure}

To demonstrate the fabrication tolerance and application in choosing the optimum waveguide structure, we use GaN as an example shown in Fig. \ref{fig: GaN_VectorField}. We show the frequency mismatch with a wider range of waveguide thickness in Fig.~\ref{fig: GaN_VectorField}(a), and the vector field of phase-matching sensitivity with respect to fabrication deviation from the target design is shown in Fig.~\ref{fig: GaN_VectorField}(b). The vector field is a representation of $\frac{\partial Mismatch(width, height)}{\partial width}$ and  $\frac{\partial Mismatch(width, height)}{\partial height}$. This serves as a direct indication of the fabrication tolerance. In areas where the magnitude of the vector is small, the phase-matching condition is less affected by changes in waveguide width and height.  A range of different dimensions can successfully achieve zero mismatch, shown by the solid black line in Fig.~\ref{fig: GaN_VectorField}(b), but the gradient plays an important role in deciding the optimum dimension that maximizes the yield. In contrast, Figs.~\ref{fig:Simulation}(b,c) show that AlGaAs and InGaP are more sensitive to waveguide dimensions, displaying a narrower range of suitable widths to achieve optimal phase matching. This sensitivity is a consequence of the Type-1 phase-matching scheme used for these materials to extend the signal-idler wavelength separation, which suggests tighter tolerances for the waveguide height and width in fabrication. 

\begin{figure}[b!]
    \centering
    \includegraphics[width=0.99\columnwidth]{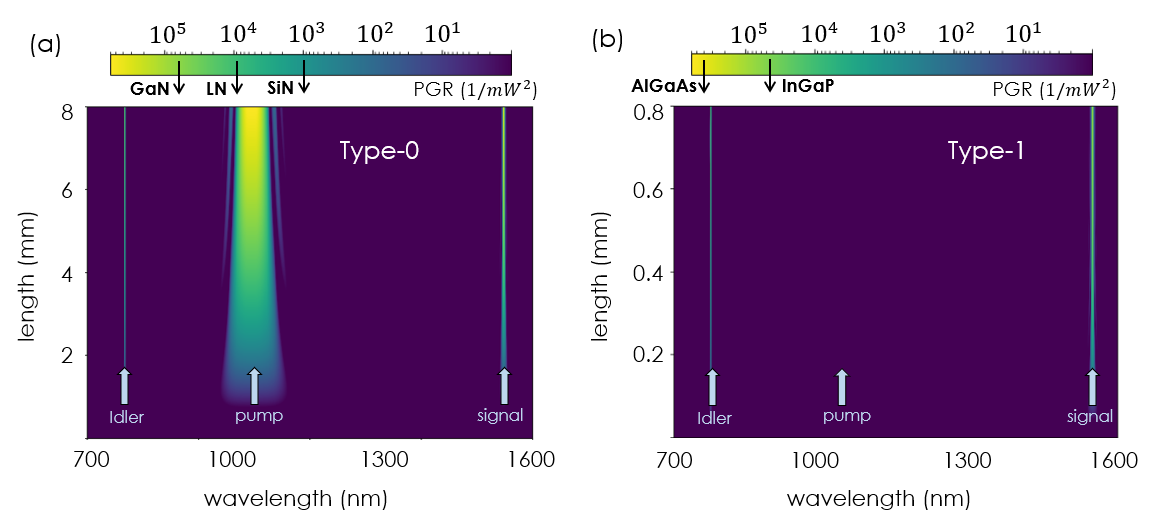}
    \caption{\label{fig: heatmap} Pair generation rate (PGR) heatmaps for Type-0 (a) and Type-1 (b) phase-matching as a function of wavelength and waveguide propagation length. The relative PGRs for Type-0 (GaN, LN, and SiN) and Type-1 (AlGaAs and InGaP) are shown on the color bar. For Type-0, both broadband pair generation as well as narrowband pair generation near the pump are possible, whereas Type-1 only allows for broadband generation due to the cross-polarized pump and signal/idler modes.}
\end{figure}

From the optimized designs with zero frequency mismatch predicted by Fig.~\ref{fig:Simulation}, we next calculate the SFWM PGR, which is shown as a function of wavelength and waveguide length (assuming a linear waveguide is used to generate entangled-photon pairs) in Fig.~\ref{fig: heatmap}. Fig. ~\ref{fig: heatmap}(a) shows the pair generation rate heatmap for GaN and employs the following equation to  determine the PGR in the low power limit for straight waveguides~\cite{helt2012does,Brainis2009,Clemmen:09}:
\begin{equation}\label{Eq:3}
    PGR_{straight} \approx \gamma^2 P^2 L_{\text{eff}}^2 \text{sinc}^2\left(\Delta k L/2\right),
\end{equation}
where $\gamma$ is the nonlinear parameter defined as $\frac{k_0 n_2}{A_{\text{eff}}}$, $P$ is the optical pump power and $L_{\text{eff}} = \frac{1 - e^{-\alpha L}}{\alpha}$.  $A_{\text{eff}}$ is the effective area, $\Delta k$ is the wavevector mismatch, $\alpha$ is the scattering loss and $n_2 = \frac{3\chi^{(3)}}{4n^2\epsilon_0 c}$ is the Kerr coefficient. Here, it is assumed that the pump is in the lower power regime.

The process for generating the plot is as follows: First, the waveguide dimensions required to achieve zero frequency mismatch are determined (see Fig.~\ref{fig: heatmap}); for instance, the optimal dimensions for GaN are approximately 650~nm by 485~nm. As previously mentioned, this cross section results in the minimized mode area for zero frequency mismatch as well as high tolerance to process variations in fabrication. Using these dimensions, the effective index of the target modes are simulated over a wavelength range from 700~nm to 1600~nm. Then $\Delta k$ is calculated as a function of wavelength and substituted into Eq.~\ref{Eq:3} to yield the pair generation rate for a straight waveguide as a function of wavelength. Note that the pump wavelength is fixed at 1033.3~nm, while we sweep the signal and idler wavelengths from 700~nm to 1600~nm while numerically calculating the pair generation rate across all wavelength combinations. The wavelength of the pump, signal, and idler photon thus must satisfy $\frac{1}{\lambda_s} + \frac{1}{\lambda_i} = \frac{2}{\lambda_p}$. The scattering loss $\alpha$ is set to 0.1 dB/cm. 

The heatmaps for SiN and LN look very similar to that of GaN but the magnitude of their PGR is different. The maximum PGR for those materials are marked on the scale bar of GaN for comparison. Fig.~\ref{fig: heatmap}(b) shows the PGR heatmap for AlGaAs, with the maximum PGR rate of InGaP marked for comparison. A key difference between the PGR behavior for GaN, LN, and SiN in Fig.~\ref{fig: heatmap}(a) and the behavior for AlGaAs and InGaP in Fig.~\ref{fig: heatmap}(b) is the SFWM signal directly adjacent to the pump near 1~$\mu$m. For materials like GaN, LN, and SiN, the effective refractive index ($n_{\text{eff}}$) remains relatively constant across the entire bandwidth. When generating entangled photon pairs near a pump wavelength of 1033~nm via SFWM, the refractive indices for signal ($n_s$), idler ($n_i$), and pump ($n_p$) photons are nearly identical, leading to minimal wavevector mismatch and maximal PGR as calculated through Eq.~\ref{Eq:3}, resulting in pair generation near the pump in addition to the broadband pairs at 775~nm and 1550~nm. In contrast, for AlGaAs and InGaP, which utilize Type-1 generation, the refractive indices for signal and idler photons differ markedly from the pump photon, primarily because $n_s$ and $n_i$ stem from the fundamental TE mode, while $n_p$ is derived from the TM mode, resulting in a larger wavevector mismatch.

Furthermore, the bandwidth near the signal and idler modes for AlGaAs and InGaP becomes progressively narrower as the waveguide length increases. This narrowing is a result of the accumulation of phase mismatch when the wavevector mismatch is nonzero. For short waveguides, the accumulated phase is small, but as the length increases, this nonzero wavevector mismatch causes the modes to become out of phase, limiting the wavelength range over which entangled photon pairs can be efficiently generated. One conclusion from this is that AlGaAs maintains a competitive PGR compared to GaN; however, the tradeoff is fabrication tolerance: achieving the target cross section to achieve high PGR is more challenging in AlGaAs and InGaP, as precise control over waveguide dimensions within a few nanometers is crucial for optimizing performance.

\begin{figure}[b!]
    \centering
    \includegraphics[width=0.99\columnwidth]{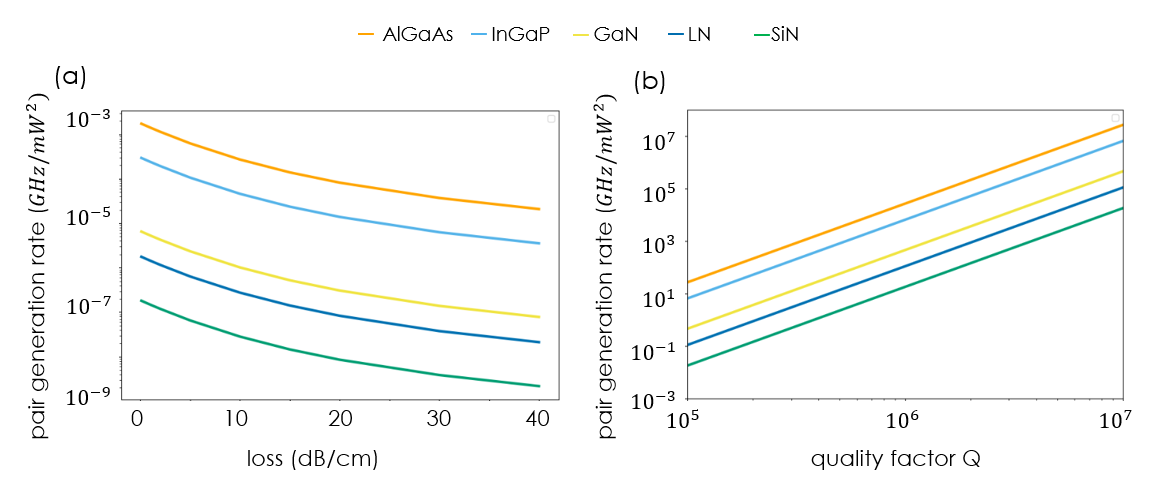}
    \caption{\label{fig:comparison}Comparison of pair generation rate for straight waveguide (a) and ring resonators (b). Note that for ring resonators, we assume $Q_i$, $Q_p$, and $Q_s$ are identical.}
\end{figure}

The comparison of PGR of different platforms in both straight waveguides and ring resonators is plotted in Fig.~\ref{fig:comparison}. For Fig.~\ref{fig:comparison}(a), we assume zero frequency mismatch and set the length of the waveguide to be 10~mm. In the case of SiN, the experimentally attainable loss for this optimized cross section is as low as 1 dB/cm~\cite{Kirill}, which corresponds to $10^{-7}$ GHz/mW$^2$ for a straight waveguide. Conversely, for AlGaAs and InGaP, although these materials experience higher loss $>1$ dB/cm for these geometries, their PGRs are still significantly greater than that of SiN primarily due to the higher nonlinearity and larger refractive indices. 

For ring resonators, the PGR is calculated using the following expression~\cite{helt2010spontaneous,azzini2012ultra}:
\begin{equation}\label{Eq:4}
    PGR_{ring} = \left( \gamma 2\pi R \right)^2 \left( \frac{Qv_g}{\omega_p \pi R} \right)^3 \frac{v_g}{4\pi R} P_{pump}^2,
\end{equation}
where $R$ is the resonator radius, $Q$ is the loaded quality factor, $v_g$ is the group velocity, and $P_{pump}$ is the optical pump power in the resonator. In the numerical simulations in Fig.~\ref{fig:comparison}(b), we set $R = 40$ $\mu$m, and $Q_s$, $Q_p$, and $Q_i$ equal to each other. From the figure, it is clear that InGaP and AlGaAs provide the highest PGR for a given $Q$. As an example, reported $Q$'s for SiN are on the order of $5 \times 10^5$ for this geometry~\cite{lu2019chip}, which yields a pair generation rate of 0.1 GHz/mW$^2$. For InGaP, $Q$'s above $4\times 10^5$ have been demonstrated at 1550 nm, which would yield a pair generation rate near 100 GHz/mW$^2$~\cite{akin2024ingap,thiel2024}. For AlGaAs~\cite{Steiner2021,steiner2023continuous} and GaN~\cite{zeng2024quantum}, which have demonstrated $Q$'s at 1550~nm greater than $10^6$, the expected PGR is also on the order of 100~GHz/mW$^2$. It is important to note that unlike the straight waveguides for which the total pair generation rate integrated over the signal and idler bandwidth is shown in Fig.~\ref{fig:comparison}(a), for resonators, the resonant enhancement increases the PGR as well as the spectral brightness, since the pairs are generated in frequency bins near 775~nm and 1550~nm with bandwidth set by the quality factor for each mode.

Although AlGaAs and InGaP demonstrate a distinct advantage in terms of PGR over other materials in both straight waveguides and ring resonators, these figures are predicated on achieving perfect phase matching; however, in practice, achieving such precision is challenging due to limitations in fabrication accuracy. Even slight dimensional variations in AlGaAs and InGaP can cause deviations from ideal phase mismatch. To compensate for these imperfections, thermal tuning may be employed to adjust and improve phase matching after the fabrication process. Additionally, wafer-scale fabrication allows for sweeping parameters to ensure full coverage of the zero frequency mismatch parameter space for a given waveguide thickness~\cite{thiel2024}. Conversely, materials like GaN and LN are not as sensitive to such process variations. They not only achieve relatively high PGRs but also meet the phase matching conditions more readily, which makes them particularly attractive for practical applications where fabrication precision may not be as stringent. Additionally, they will likely exhibit lower propagation loss at 775~nm, and ultimately exhibit broader bandwidth for higher PGRs deeper in the visible regime.


\section{Conclusion}

In this study, we compared different photonic material platforms for generating entangled photon pairs via spontaneous four-wave mixing to bridge the gap between long-distance communication at telecom wavelengths and visible-wavelength quantum sensors, processors, and memories. Through detailed numerical simulations, we found that materials like AlGaAs and InGaP can achieve the highest pair generation rates for a given resonator quality factor $Q$, but they also have the most stringent requirements on the fabrication precision required to achieve optimal phase matching. Through a Type-1 cross-polarized phase-matching scheme, their idler wavelength can also be extended deeper into the visible regime. Other materials such as silicon nitride have a lower pair generation rate, but they can also benefit from industrial-standardized foundry manufacturing processes and exhibit a higher tolerance to fabrication process variations due to flat dispersion. Of all the materials studied, GaN may represent the optimal choice in that it exhibits a moderately large $\chi^{\left(3\right)}$ nonlinearity, less sensitivity to material parameters and process variations, and thus phase matching is relatively straightforward; however, low-loss GaN fabrication and processing, including various alloys such as AlGaN, AlN, and ScAlN, is an emerging direction for exciting future studies for nonlinear photonics and broadband pair generation~\cite{liu2023aluminum,xu2024silicon}.

Another important consideration is the linewidth of the signal and idler photons. For applications in quantum networking and distributed computing, the idler photons may interact with a variety of quantum systems with optical transitions in the visible regime, such as those indicated in Fig.~\ref{fig:intro}. This interaction may be done through resonant linear transmission or reflection from the quantum systems or through two-photon interference, such as Bell-state measurements for entanglement swapping and teleportation. In these cases, not only is the resonance wavelength of the idler photon important, but for high-fidelity and high-efficiency entanglement swapping, the spatial, spectral, and temporal modes of the idler photon and the photon emitted from the quantum system must also match. For example, the zero-phonon linewidths of color centers in diamond and silicon carbide~\cite{PRXQuantum.1.020102} can be as low as $\sim10$~MHz. Generating an idler photon with a similar linewidth would require $Q$ on the order of 40~$\times 10^6$, which is attainable with SiN. On the other hand, hBN quantum defects with spin-photon transitions and semiconductor QDs have zero-phonon linewidths ranging from $\sim100-500$ MHz. These linewidths correspond to $Q$s on the order of $10^5-10^6$, which is attainable with each of the studied material platforms~\cite{moody2020chip,azzam2021prospects}.

\section*{Data Availability Statement}

\noindent All data that supports the findings of this study are available upon request.

\begin{acknowledgement}

This work was supported by the NSF Quantum Foundry through the Q-AMASE-i Program (Grant No. DMR-1906325), the NSF CAREER Program (Grant No. 2045246), the Keysight University Research Program, and the Eddleman Center for Quantum Innovation. L. T. acknowledges support from the NSF Graduate Research Fellowship Program.

\end{acknowledgement}




\bibliography{moodybib}

\end{document}